\documentclass[journal=jacsat,manuscript=article]{achemso}

\usepackage[version=3]{mhchem} % Formula subscripts using \ce{}

\author{Atharva Paranjape}

\affiliation[Max Planck Institute for Solid state Research 70569, Stuttgart, Germany]
{Max Planck Institute for Solid State Research, 70569, Stuttgart, Germany}

\author{Kathrin Küster}
\affiliation[Max Planck Institute for Solid state Research 70569, Stuttgart, Germany]
{Max Planck Institute for Solid State Research, 70569, Stuttgart, Germany}

\author{Olga Shevstova}
\affiliation[Max Planck Institute for Solid state Research 70569, Stuttgart, Germany]
{Max Planck Institute for Solid State Research, 70569, Stuttgart, Germany}

\author{Lisa Ebo}
\affiliation[Max Planck Institute for Solid state Research 70569, Stuttgart, Germany]
{Max Planck Institute for Solid State Research, 70569, Stuttgart, Germany}

\author{Toni Hache}
\affiliation[Max Planck Institute for Solid state Research 70569, Stuttgart, Germany]
{Max Planck Institute for Solid State Research, 70569, Stuttgart, Germany}

\author{Klaus Kern}
\affiliation[Max Planck Institute for Solid state Research 70569, Stuttgart, Germany]
{Max Planck Institute for Solid State Research, 70569, Stuttgart, Germany}

\author{Rainer Stöhr}
\affiliation[University of Stuttgart]
{3\textsuperscript{rd} Institute of Physics and Research Center SCoPE, University of Stuttgart, 70049, Stuttgart, Germany}

\author{Jörg Wrachtrup}
\affiliation[University of Stuttgart]
{3\textsuperscript{rd} Institute of Physics and Research Center SCoPE, University of Stuttgart, 70049, Stuttgart, Germany}

\author{Aparajita Singha}
\email{aparajita.singha@tu-dresden.de}
\affiliation[Max Planck Institute for Solid state Research 70569, Stuttgart, Germany]
{Institute of Solid State and Materials Physics, Dresden University of Technology, 01069, Dresden, Germany}
\alsoaffiliation{Würzburg-Dresden Cluster of Excellence ctd.qmat, 01069, Dresden, Germany}
\alsoaffiliation{Max Planck Institute for Solid State Research, 70569, Stuttgart, Germany}

\title[An \textsf{achemso} demo]
  {Electrostatic Stabilization of Near-Surface Quantum Sensors via Dielectric Interface Engineering}
\abbreviations{IR,NMR,UV}

\begin{document}

%\begin{tocentry}

%Some journals require a graphical entry for the Table of Contents.
%This should be laid out ``print ready'' so that the sizing of the

%Inside the \texttt{tocentry} environment, the font used is Helvetica
%8\,pt, as required by \emph{Journal of the American Chemical

%The surrounding frame is 9\,cm by 3.5\,cm, which is the maximum
%permitted for  \emph{Journal of the American Chemical Society}
%graphical table of content entries. The box will not resize if the
%content is too big: instead it will overflow the edge of the box.

%This box and the associated title will always be printed on a
%separate page at the end of the document.

%\end{tocentry}

\begin{abstract}
 Control of charge-state stability in near-surface quantum defects is critical for nanoscale sensing, yet remains particularly challenging under ultra-high vacuum (UHV), where surface-induced band bending destabilizes the metrologically relevant charge-state. Here, we present a robust and reproducible approach for stabilizing shallowly implanted ($<$ 10 nm deep) nitrogen-vacancy (NV) centers  in near-UHV conditions (P = 3 × 10\textsuperscript{-9} mbar)  based on dielectric interface engineering. Through measurements on individually addressable NV centers, we demonstrate that a TiO\textsubscript{2} coating on the diamond suppresses surface-induced electrostatic fields, yielding a $79\%$
 increase in NV\textsuperscript{-} population and a $46\%$ enhancement in NV-spin resonance contrast at room temperature. Coherent control measurements further reveal suppressed charge-state conversion dynamics. These results establish dielectric screening as a powerful and reproducible strategy to engineer charge transition energetics of NV centers in scanning-probe-compatible geometries under extreme conditions.
 
 %% Older version 
 %Nitrogen vacancy (NV) centers in diamond constitute a promising platform for highly sensitive measurements of magnetic fields at the nanoscale. In the well-controlled isolated environment of ultra-high vacuum (UHV), near-surface NV centers could enable sensing and manipulation of proximal spins. However, shallowly implanted NV-centers are highly prone to charge-state instabilities under such experimental conditions. Here, we use UHV compatible surface treatments to modify the local environment of NV centers located few nanometers from the diamond surface. Using single, individually addressable NV centers hosted in nanostructured diamond, we demonstrate with statistical significance that controlled coating of the diamond surface with titanium dioxide (TiO\textsubscript{2}) drastically enhances the charge-state stability of shallow NV centers, characterized by $74\%$ increase in NV- population, $45\%$ increase in spin contrast and $21\%$ improvement in magnetic field sensitivity at ambient temperature and P = 3 x 10\textsuperscript{-9} mbar). These results establish a reproducible approach to utilize shallow NV sensors in scanning probe compatible geometry under extreme conditions albeit without compromising on their optical and spin properties. 
\end{abstract}


Building on pioneering proof-of-principle experiments, \cite{Gruber1997ScanningCenters,Balasubramanian2008NanoscaleConditions,Maze2008NanoscaleDiamond,Taylor2008High-sensitivityResolution} nitrogen-vacancy (NV) centers in diamond have evolved into a widespread and powerful quantum platform capable of sensing magnetic fields with extreme precision~\cite{Rondin2014MagnetometryDiamond,Degen2017QuantumSensing}. Besides magnetic field sensing, NV centers have now been utilized for measurements of temperature \cite{Neumann2013High-PrecisionDiamond}, electric fields \cite{Dolde2011Electric-fieldSpins}, and nanoscale NMR measurements \cite{Mamin2013NanoscaleSensor}. The workhorse for these applications is the negatively charged NV\textsuperscript{-}, owing to spin-dependent photoluminescence (PL), optical state initialization and coherent spin state control. Near-surface NV centers which can be brought in close proximity to systems of interest have widespread applications ranging from surface NMR \cite{Liu2022SurfaceDiamond} to characterization of nanoscale devices~\cite{Hache2025NanoscaleSensor,Ku2020ImagingGraphene} to modern multiplexed scanning NV magnetometry approach \cite{Huxter2025MultiplexedSensors}. Despite this, such shallow NV centers are strongly affected by charge-state instability originating from the presence of charge traps and defects near the surface \cite{Janitz2022DiamondCenters}.  Particularly in scanning NV magnetometry, this impacts the achievable spatial resolution, which improves as the sensor is closer to the source of magnetic information and thus closer to the diamond surface \cite{QnamiAG2021TechnicalMagnetometry}. Current nanoscale sensing experiments are therefore limited by a trade-off between maximizing spatial resolution and maintaining charge-state stability. Generation of charge-stable NV centers is thus crucial to address this challenge and enable high resolution, quantitative and non-invasive quantum magnetometers.\

The routine operation of such quantum magnetometers in  ultra-high vacuum (UHV) is currently hampered by  severe charge-instabilties \cite{Neethirajan2023ControlledCenters}. However, UHV offers a pristine, extremely controlled and isolated environment, opening up several avenues for highly precise experiments involving probing and manipulating external spins at the individual or few molecule level \cite{Schlipf2017AQubit,Pinto2020ReadoutSpin, Wu2020ImagingGlycans, Anggara2020ExploringCollision}. Previous works on charge-state stabilization using surface treatments have shown improvements on single NV centers in ambient conditions \cite{Zhang2019EnhancingLayers,Giri2023ChargeDiamond,Bian2021NanoscaleCondition} as well as  NV ensembles in vacuum (10$^{-3}$ mbar) \cite{Kumar2024StabilityPassivation}. Other than surface treatments, stabilization in ambient conditions by utilizing high power laser exposure for ensemble NV centers \cite{Gorrini2021Long-LivedPhotoconversion} and novel diamond growth methods to stabilize NV centers \cite{Kageura2022ChargeDiamond,Alkahtani2020ChargeNanodiamonds} have also been employed. Although methods for achieving charge-state stability under UHV conditions have recently been explored via physisorption of well defined molecules \cite{Neethirajan2023ControlledCenters}, a robust chemistry-based approach to preserve the NV\textsuperscript{-} state in these extreme conditions has yet to be established.\

Here, we address this challenge by employing a repeatable, homogeneous, chemically stable, and vacuum compatible surface treatment for enhancing charge-state stability of shallowly implanted NV centers in a clean environment. Crucially, we investigate charge-state stability of individual NV centers in tip-shaped diamond nanopillars, enabling direct knowledge transfer to scanning probe geometries operating under extreme conditions. We report that the deposition of a TiO\textsubscript{2} coating on the diamond surface results in (a) an average of 46 $\%$ enhancement of contrast in Optically Detected Magnetic Resonance (ODMR) spectroscopy, (b) an average of 79 $\%$ increase of NV\textsuperscript{-} contribution observed in the fluorescence spectrum, %(c) 21 $\%$ increase in DC magnetic field sensitivity
based on measurements on 17 different single NV centers. Through these measurements, we establish a reproducible route to obtain charge-stable shallow NV centers in UHV.  \

\begin{figure*}[h]
    \centering
    \includegraphics[width=\columnwidth]{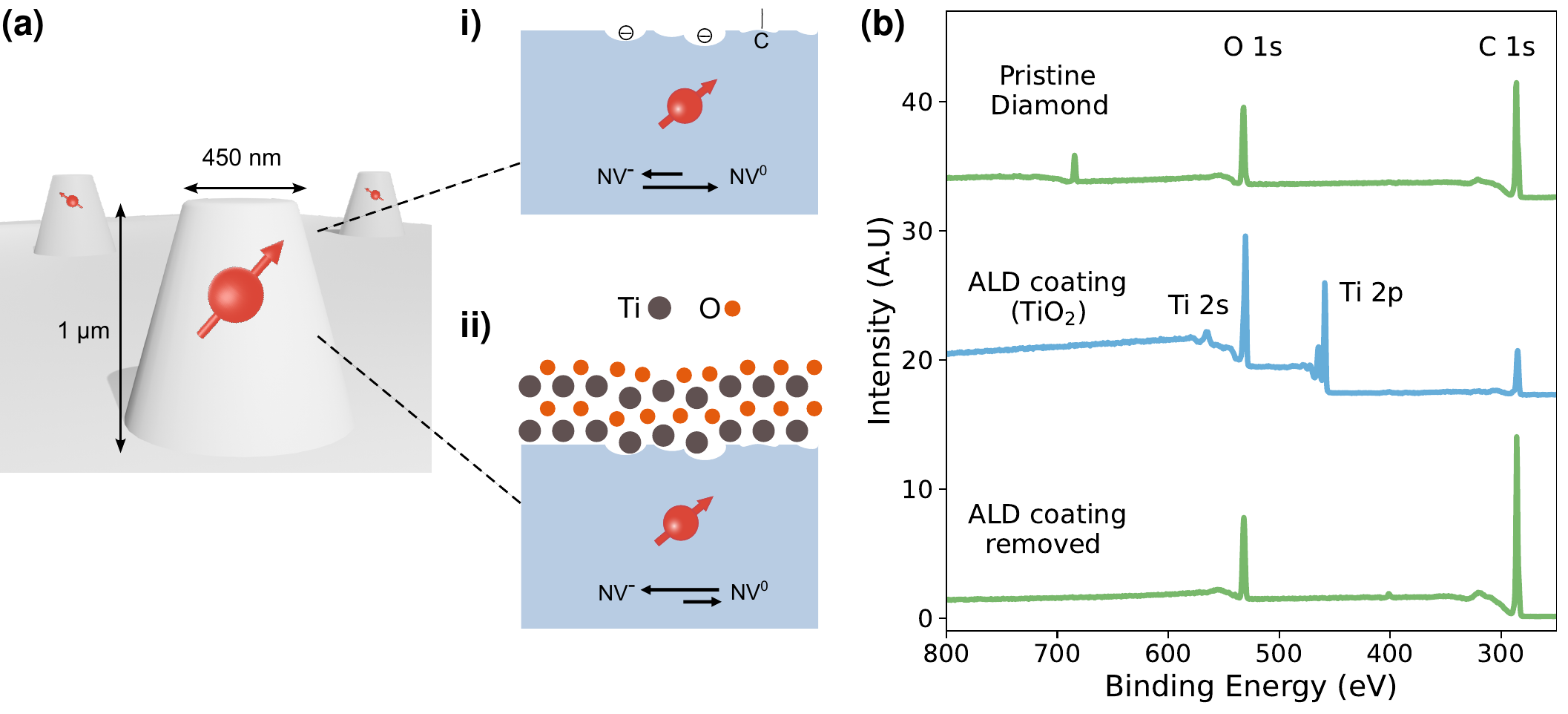}
    
    \captionof{figure}{ \textbf{Schematic of dielectric coated diamond and surface characterization}  (a) Electronic environment of NV centers in nano-structured diamond depicted under i) pristine and ii) TiO$_{2}$ coated conditions. The length of black arrows represents the relative strength of charge-state switching. b) X-ray Photoelectron Spectroscopy of a reference sample (RS) under different surface conditions. Ti signal is observed upon ALD deposition (coating performed using 500 cycles of ALD). These signatures are removed after cleaning the diamond surface. A residual peak at the top panel at 684.8 eV corresponds to F 1s.}
    \label{fig:setup}
\end{figure*}

The schematic of the diamond membrane sample employed in the experiments is shown in Figure \ref{fig:setup}(a). The NV centers are hosted in an array of nanopillars etched on the diamond \cite{Momenzadeh2015NanoengineeredCenters}, nominally containing one NV center per pillar. Crucially for near-surface NV centers, the presence of surface roughness and unsaturated bonds at the surface can create trapping potentials for charges \cite{Janitz2022DiamondCenters} and thereby influence the local electronic environment of the defects as depicted in Figure \ref{fig:setup} (a,i). The local environment directly affects the charge-state dynamics of the NV centers, and by extension, the performance of NV-based sensing techniques.  Hence a modification of the surface is expected to affect these dynamics and provides a tool to control the charge-state \cite{Zhang2019EnhancingLayers,Giri2023ChargeDiamond,Kumar2024StabilityPassivation,Neethirajan2023ControlledCenters,Rondin2010Surface-inducedNanodiamonds}.

%The implanted NV centers can be found in two different fluorescent charge-states, the negatively charged NV\textsuperscript{-} and the neutral NV\textsuperscript{0}, which can be distinguished by their characteristic fluorescence spectra. 
%The NV\textsuperscript{-} has a zero-phonon-line (ZPL) around 637 nm  while the NV\textsuperscript{0} has a ZPL around 575 nm \cite{Davies1976OpticalDiamond, Doherty2013TheDiamond}. Under 515 nm laser illumination, the NV center undergoes continuous charge-state switching \cite{Aslam2013Photo-inducedDetection,Gaebel2006PhotochromismDiamond,Shields2015EfficientConversion}. Consequently, any time-averaged measurement consists of contributions from both charge-states. For fixed illumination and measurement parameters, the observed charge-state equilibrium of individual NV centers is then determined by their individual nanoscale local environment. Surface roughness and unsaturated bonds at the surface can create trapping potentials for charges \cite{Janitz2022DiamondCenters}. This electronic environment as shown in Figure \ref{fig:setup} (a,i) thus plays a role in determining the charge-state dynamics of the NV centers. Hence a modification of the surface is expected to affect these dynamics and provides a tool to control the charge-state as shown in previous works\cite{Zhang2019EnhancingLayers,Giri2023ChargeDiamond,Kumar2024StabilityPassivation,Neethirajan2023ControlledCenters,Rondin2010Surface-inducedNanodiamonds} .\

In this work, in order to perform a chemically stable and repeatable controlled surface treatment, we deposit thin layers of TiO\textsubscript{2} on the diamond surface using atomic layer deposition (ALD) as shown schematically in Figure \ref{fig:setup} (a,ii). TiO\textsubscript{2} is a dielectric material which is transparent in the visible spectrum of light \cite{Diebold2003TheDioxide} and can be deposited with high precision \cite{Xie2007AtomicH2O} without introducing large scale magnetic impurities. Hence it is expected not to interfere with NV magnetometry protocols involving optical readout of the NV spin. In addition, note that our chosen TiO\textsubscript{2} thickness of 6 nm is significantly smaller than the emission wavelength of the NV\textsuperscript{-} center (637 nm+). This, together with minimal refractive index mismatch of this thin film dielectric with respect to that of diamond, rules out spurious optical artifacts in our measurements such as modified collection efficiency, interference effects and/or cavity effects. Furthermore, TiO\textsubscript{2} is chemically robust under contact with neutral and basic environments \cite{Correa2015ChemicalDeposition} without the need for annealing\cite{Broas2017ChemicallyProcessability}. Thus TiO\textsubscript{2} ALD provides a simple, repeatable,  coating technique, well suited for contact with a wide range of samples of interest \cite{Liu2022SurfaceDiamond}. \

We begin our study by establishing the successful application and subsequent removal of the coating from the nanostructured diamond sample. For this, we deposit TiO\textsubscript{2} using 500 cycles of ALD (150 °C) on a reference diamond sample (RS) with etched nanopillar arrays, however without implanted NV centers. An X-Ray Photoelectron Spectroscopy (XPS) measurement of the sample before and after ALD deposition is shown in the upper and middle panel of Figure \ref{fig:setup} (b) respectively. The residual peak in the upper panel at 684.8 eV corresponding to F 1s results from imperfect cleaning during reference sample preparation. The observation of Ti 2s and Ti 2p peak in the middle panel clearly demonstrates deposition of TiO\textsubscript{2} on the nanostructured diamond. Subsequently the RS is cleaned using Piranha solution for three hours. The absence of the Ti peaks in the corresponding XPS measurements of the cleaned RS (lower panel of Figure \ref{fig:setup} (b)) validates that the coating is entirely removed by this cleaning process, effectively returning the diamond to pristine condition. This versatility offered by the straightforward application and removal of the coating makes the treatment particularly attractive as the same sample can be reconfigured for subsequent surface treatments.

To investigate the impact of the TiO\textsubscript{2} coating on the optical and spin properties of the sensors, we now turn to the primary sample with NV centers shallowly implanted within nanostructured diamond. All the subsequent measurements reported in the Letter are from this primary sample without explicitly mentioning the same. The sample is transferred to the UHV sample stage, and all subsequently reported measurements are carried out at a pressure of P = 3 x 10\textsuperscript{-9} mbar and temperature T=300 K (see Methods). A region of the sample implanted with \textsuperscript{15}N ions with 5 keV implantation energy is used for the measurements. The mean depth of the implanted ions is 8 $ \pm$ 3 nm as calculated using SRIM (Stopping and range of ions in matter) simulation \cite{Neethirajan2023ControlledCenters, Ziegler2010SRIM2010}. The NV centers are excited using a 515 nm laser and the fluorescence is collected using a home-built confocal microscope (see SI Figure 7).
To characterize the charge-state properties, multiple single shallow NV centers hosted in independent nanopillars are first selected in the pristine diamond/uncoated condition, which are then followed for all subsequent measurements both before and after the ALD coating. The presence of exactly one NV center in a nanopillar is verified prior to any experiment reported here, via autocorrelation measurements (see SI Figure 6) by guiding the readout optical signal through a 650 nm long-pass (LP) filter. In the following sections, we present a detailed characterization of the charge-state and spin properties of the NV centers as impacted by the surface treatment. \

For direct information on the NV charge-state, we measure the fluorescence spectrum of the selected single NV centers. The fluorescence spectra are measured using a separate detection path after passing the signal through a 550 nm longpass filter (see SI Figure 7). Subsequently, we deposit TiO\textsubscript{2} on the membrane sample using 100 cycles (150 °C) of ALD. The thickness of the layer is estimated using spectroscopic ellipsometry approach on a co-deposited silicon reference sample to be around 6.07 $\pm$ 0.38 nm. Then the charge-state of the same selected set of NV centers is investigated in the coated sample by measuring the fluorescence spectra. \

\begin{figure*}[!ht]
\centering
\includegraphics[width=\textwidth,
    height=0.75\textheight,
    keepaspectratio]{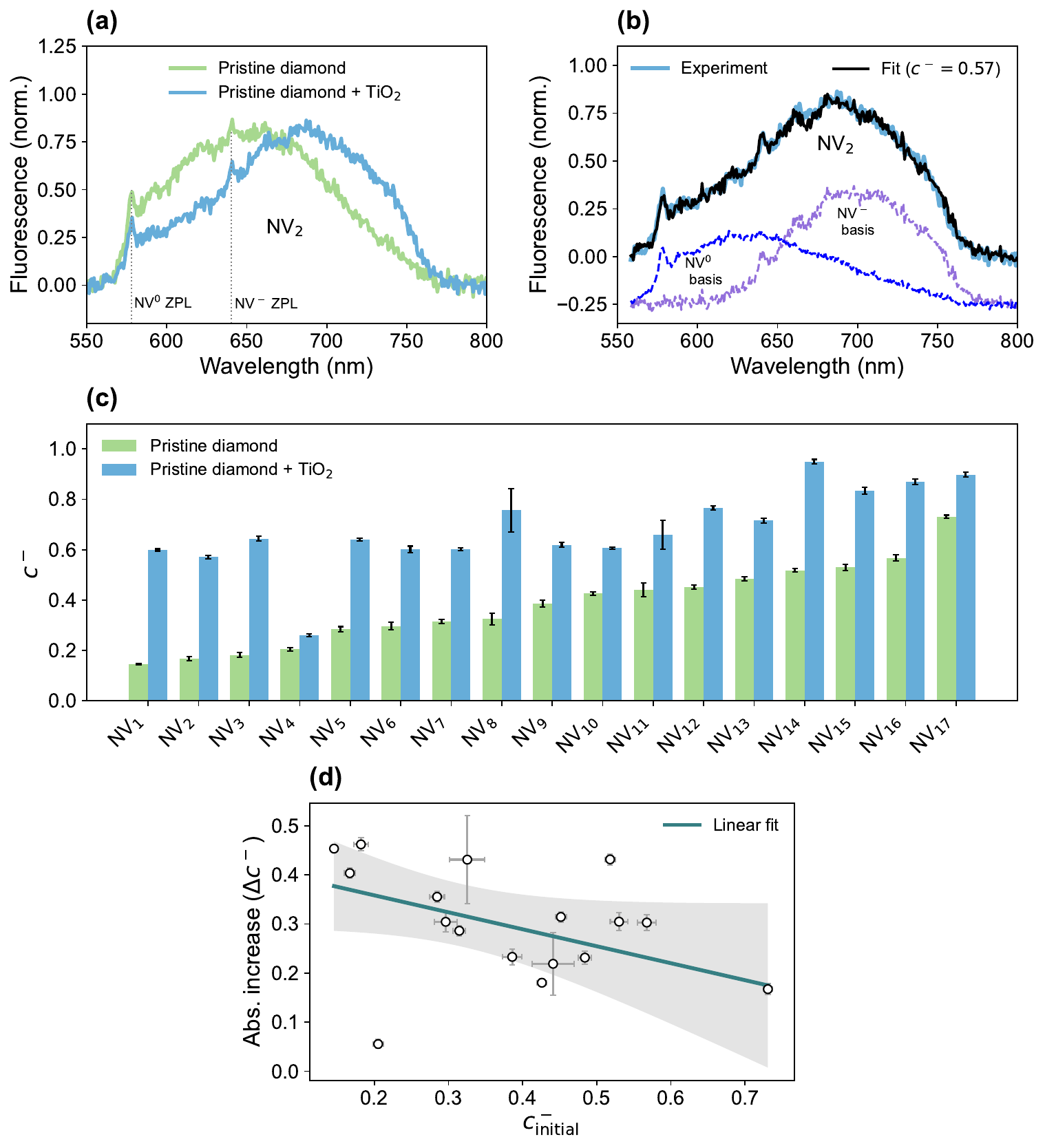}
\caption{\textbf{Fluorescence spectrum of individual NV centers in near-UHV condition} (a)
Measured fluorescence spectrum of the same single NV center (NV\textsubscript{2}) under pristine and coated conditions. Plotted spectra are area-normalized. P= 3 x 10\textsuperscript{-9} mbar, T= 300 K, incident laser power before the objective = 130 $\mu$W. (b) Decomposition of measured spectrum using pure NV\textsuperscript{-} and NV\textsuperscript{0} spectra as basis functions to obtain NV\textsuperscript{-} percentage. The least squares fit (black) is in excellent agreement with the experimentally obtained spectrum (purple) and we obtain  \(c^{-}\)= 0.57. (c) Statistics of \(c^{-}\) in pristine and coated diamond conditions (d) Improvement in \(c^{-}\)  after ALD coating as a function of initial c\textsuperscript{-}. Linear fit (blue) is shown with the 95$\%$ confidence interval (grey) for the fit. } 
\label{fig:spectrum} 
\end{figure*}

In Figure \ref{fig:spectrum} (a), we present the comparison of fluorescence spectra from the same representative single NV center (NV\textsubscript{2}) before and after coating the diamond. The plotted data is area-normalized to have integrated area = 100. While the peaks of the ZPL from both NV\textsuperscript{-} and NV\textsuperscript{0} are observed, a clear shift in the peak of the phonon sideband towards 690 nm is visible after the surface treatment. Under 515 nm laser illumination, the NV center undergoes continuous charge-state switching \cite{Aslam2013Photo-inducedDetection,Gaebel2006PhotochromismDiamond,Shields2015EfficientConversion}. The two fluorescent charge states exhibit a distinct spectral signature, with the NV\textsuperscript{-} having a zero-phonon-line (ZPL) around 637 nm  while the NV\textsuperscript{0} having a ZPL around 575 nm \cite{Davies1976OpticalDiamond, Doherty2013TheDiamond}. Furthermore, the peak of the phonon-side-band also changes from 700 nm for the NV\textsuperscript{-} to 625 nm for the NV\textsuperscript{0}. Due to continuous charge-state switching, any time-averaged measurement consists of contributions from both charge-states. For fixed illumination and measurement parameters, the observed contribution in optical measurements from each charge-state provides a proxy for the  charge-state equilibrium of individual NV centers. Therefore, the shift in the peak of the phonon side band towards 690 nm as shown in  \ref{fig:spectrum} (a) represents a higher dwell time in the NV\textsuperscript{-} charge-state. These results demonstrate that the controlled surface modification provides direct access to the charge-state of the NV, seen via a shift in the charge-state equilibrium.

For a quantitative analysis, the experimentally obtained spectrum is decomposed into contributions from  NV\textsuperscript{-} and NV\textsuperscript{0} charge-states similar to the method used by Alsid et al \cite{Alsid2019PhotoluminescenceDiamond} (see SI Figure 1 for details). For this purpose, the area-normalized experimentally recorded spectrum is represented as: 
\begin{equation}
     I(\lambda)_{norm} \ = \ c^{0}  \times  \hat{I}_{0}(\lambda) \\ + \\ c^{-} \times \hat{I}_{-}(\lambda)
     \label{eq:decomp}
\end{equation}
Here, \(c^{0}\) and \(c^{-}\) represent the fractional contribution to the fluorescence, and $\hat{I}_{0}(\lambda)$ and $\hat{I}_{-}(\lambda)$ represent the area-normalized spectral basis functions for NV\textsuperscript{0} and NV\textsuperscript{-} respectively. The basis functions are extracted using extreme cases of charge-states from the same sample from independent measurement rounds. Considering the  NV\textsuperscript{-} and  NV\textsuperscript{0} as the only contributing fluorescent charge-states, we carry out a least squares fit using equation (\ref{eq:decomp}) with the constraint \(c^{-}\) + \(c^{0}\) = 1, and  \(c^{-}\) as the free parameter. The spectral decomposition in the case of the same single NV center (NV\textsubscript{2}) in TiO\textsubscript{2} coated diamond is shown in Figure \ref{fig:spectrum} (b). The fit is in excellent agreement with the experimentally obtained spectrum and reveals a \(c^{-}\) of 0.57. Hence we find for NV\textsubscript{2}, the contribution increases significantly from \(c^{-}\)=0.16 to \(c^{-}\) 0.57, corresponding to a 3.4 fold enhancement . We note that c\textsuperscript{-} represents the fractional contribution of NV\textsuperscript{-} to the total fluorescence measured and may not reflect the exact NV\textsuperscript{-} population, as the fluorescence output from the two charge-states may differ appreciably. However, as the NV\textsuperscript{-}/NV\textsuperscript{0} flourescence factor $\kappa$ is greater than 1 , the enhancement factor calculated using c\textsuperscript{-} is only the lower bound, and the enhancement factor for the NV\textsuperscript{-} population is potentially higher (Refer SI Figure 1 for details).

This decomposition analysis is extended to all 17 NV centers, both before and after the surface modification, revealing the striking effect on the charge-state stability.  Figure \ref{fig:spectrum} (c) shows the statistics of the  NV\textsuperscript{-} fractional contribution (\(c^{-}\)) of individual NV centers.  We report that the average \(c^{-}\) from the 17 selected NV centers increases from 0.38 to 0.68 as a direct consequence of the surface treatment. This represents an average of 79$\%$ increase in the NV\textsuperscript{-} contribution to the fluorescence. With this significant improvement, we note that the coating acts as a global surface treatment which simultaneously improves NV centers hosted in separate nanopillars, which can be employed during scanning NV tip fabrication and/or membrane sample design. Additionally we note that the effect of the coating is reversible (refer SI Figure 4 ). This provides evidence that it is indeed the surface treatment that shifts the charge-state equilibirum, providing the benefit of a reconfigurable surface for further tailoring of surface properties.

%This demonstrates that the TiO\textsubscript{2} coating stabilizes the charge state of the NV centers towards NV\textsuperscript{-} in UHV. 

To further understand the effect of the coating, we look at the improvement in \(c^{-}\) defined as:

\begin{equation}
   \Delta C =  [c^{-}_{final}-c^{-}_{initial}] \times 100
\end{equation} 

where \(c^{-}_{intial}\) and  \(c^{-}_{final}\) represent the value of \(c^{-}\) obtained from measurement on the same NV center before and after the ALD coating. This improvement is shown as a function of the \(c^{-}_{initial }\) of each NV in Figure \ref{fig:spectrum} (d). We find that the increase in the \(c^{-}\) is strongly correlated with the initial condition. The increase is the highest for the NV centers with low \(c^{-}_{initial }\) contribution and is the least for the NV centers with high \(c^{-}_{initial}\) contribution. This effect can be explained by considering the different environments that each NV center is surrounded by. When the \(c^{-}_{intial}\) is low, the dominance of the NV\(^0\) charge-state indicates the presence of a large number or strongly disruptive charge traps. When the diamond is coated with the TiO\textsubscript{2}, these traps are likely passivated leading to a significant enhancement in the \(c^{-}\). On the other hand, when the \(c^{-}_{intial}\) is high, the local environment of the NV can be understood to be relatively defect free such as in bulk diamond. Hence the passivation has minimal impact on these NV centers. This behavior indicates that using the ALD coating, we are able to induce a change in the local electronic environment of the shallow NV centers. This change is then manifested as the shift in the charge-state equilibrium.  \

\begin{figure}[!h]
\centering
\includegraphics[width=\linewidth]{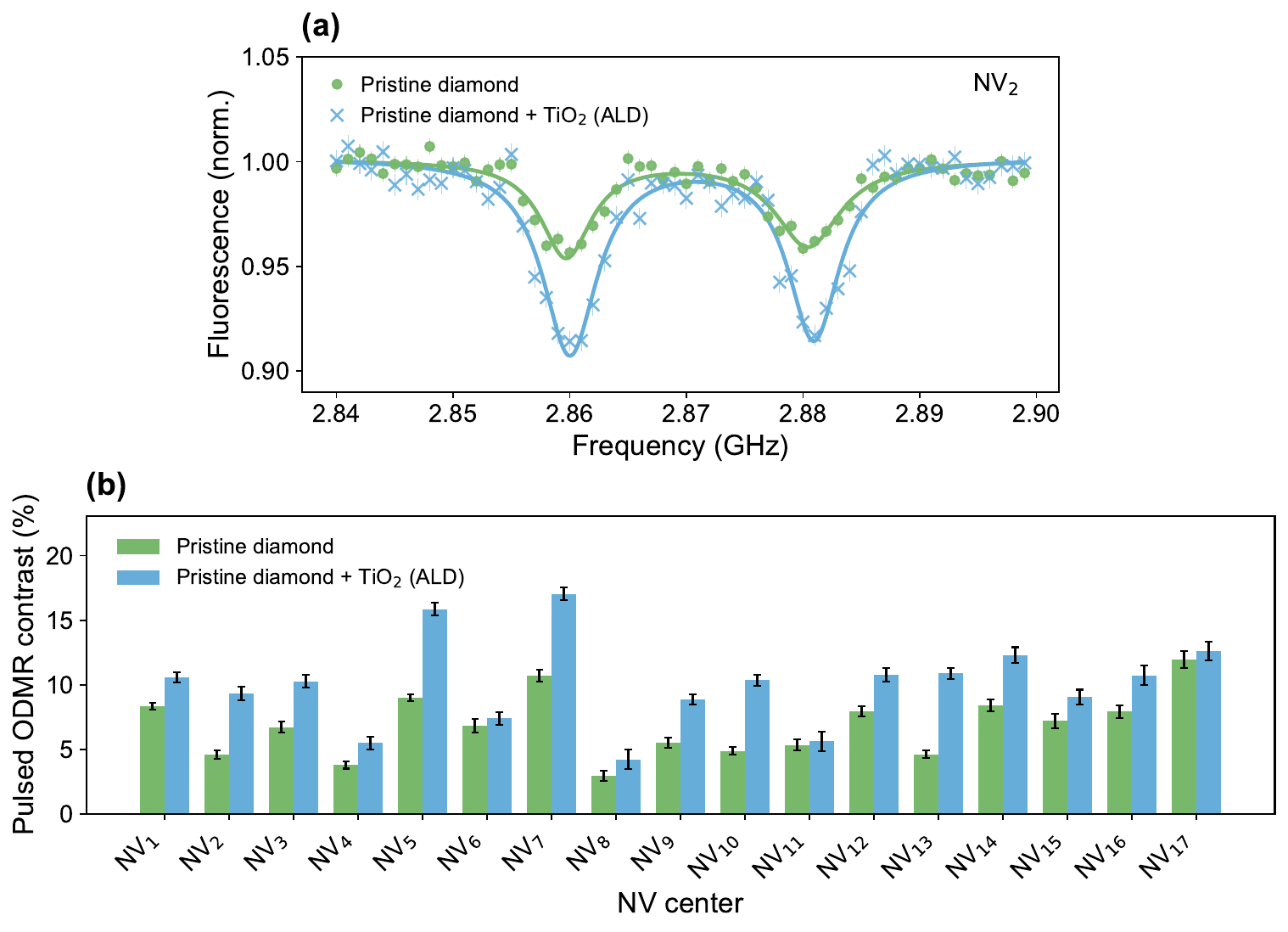}
\caption{\textbf{Pulsed ODMR spectroscopy of single NV centers in near-UHV condition} (a) Pulsed ODMR measurement of NV\textsubscript{2} (same NV as in Figure 2) before and after the coating. P= 3 x 10\textsuperscript{-9} mbar, T= 300 K, incident laser power before the objective = 130 $\mu$W  (b) Statistics of the ODMR contrast in pristine and ALD coated conditions for 17 measured NV centers.}
\label{fig: odmr}
\end{figure}

Having established the consistent and reproducible improvement in charge-state stability, we further investigate the resulting impact of the same on measurement techniques central to NV based-sensing. For this, we look at pulsed microwave and laser based sensing protocols, which offer precise control over the spin state of the NV\textsuperscript{-}. 
%In order to enable functionality in complex magnetometry techniques, it is crucial to preserve the spin properties of the NV\textsuperscript{-}. 
%With the stable NV\textsuperscript{-} charge-state, it is also important to preserve the spin properties of the NV\textsuperscript{-} in order to enable functionality in complex magnetometry techniques. 
%Pulsed microwave and laser based sensing protocols offer precise control over the spin state of the NV\textsuperscript{-}. 
Hence we investigate the effect of the TiO\textsubscript{2} coating on the pulsed ODMR spectrum. The ODMR spectrum for each NV is recorded immediately after the acquisition of the fluorescence spectrum for the corresponding NV. %The pulse sequence used for this measurement is shown in the inset of Figure \ref{fig: odmr} (a). 
An initial laser pulse polarizes the NV to the m\textsubscript{s}=0 spin sublevel, then a MW pulse centered around 2.87 GHz converts the spin state to  m\textsubscript{s}= $\pm$ 1 state. A final laser pulse is used to read out the spin state as well as to repolarize the NV to m\textsubscript{s}=0 state.

Figure \ref{fig: odmr} (a) shows the pulsed ODMR measurement performed on the same single NV center (earlier denoted as NV\textsubscript{2} in Figure \ref{fig:spectrum} ) before and after coating with TiO\textsubscript{2}. A static magnetic field of 0.4 mT lifts the degeneracy of the m\textsubscript{s}= $\pm$ 1 states leading to two dips in the NV spin resonance. The flat region of the double Lorentzian fit is normalized to 1, and the contrast is then defined as the percentage difference between the fit’s maximum and minimum values. An increase in the ODMR contrast of the same NV center from 4.6 $\pm$ 0.3 to 9.3 $\pm$ 0.5 is clearly seen after the ALD coating. This represents a remarkable increase of 102 $\%$ in the contrast. The characteristic ODMR signature of a flourescence dip centered around 2.87 GHz can be observed only in the NV\textsuperscript{-} state, while the NV\textsuperscript{0} contributes to background fluorescence \cite{Felton2008ElectronDiamond,Bluvstein2019IdentifyingCenters}. Hence, in a time-averaged ODMR measurement involving continuous charge-state switching, the ODMR contrast has a strong dependence on the dwell time in each charge state. From the increased contrast, we deduce an increase in the dwell time in NV\textsuperscript{-} hence a shift in the charge-state equilibrium. This observation is consistent with the results seen through the fluorescence spectrum. To verify the reproducibility of this phenomenon, we look at comparison of the ODMR spectra of all 17 NV centers, which remarkably reveal a consistently increased spin resonance contrast in all measured NV centers.  The comparison of the contrast of the NV centers before and after the coating is shown in Figure \ref{fig: odmr} (b). We find that the average ODMR contrast increases from 6.9 $\pm$ 0.6 $\%$ to 10.1 $\pm$ 0.8 $\%$ as a result of the surface modification, which represents an enhancement of 46$\%$. Notably, none of the NV centers exhibited reduction in the pulsed ODMR contrast. An immediate impact of the increased contrast is the improved sensitivity of the NV center to detect static magnetic fields. From these measurements, we estimate that the average  photon shot-noise limited sensitivity \cite{Rondin2014MagnetometryDiamond} improves by  19 $\%$  (see SI Figure 2 for details). %Thus we demonstrate that the controlled surface modification used in this work 

%Differing from continuous wave (CW) ODMR, pulsed ODMR involves coherent control over the NV spin. 
While pulsed ODMR enables DC magnetometry, AC sensing protocols are critically dependent on the ability to coherently manipulate the NV spin. %Previous studies have established that the NV spin coherence degrades in near-surface conditions \cite{Sangtawesin2019OriginsSpectroscopy}.
In order to understand the effect of our ALD treatment in this regard, we compared the driven Rabi oscillations of NV centers before and after the coating. %The inset of Figure \ref{fig: rabi} (a) shows the pulse sequence used for these measurements. 
An initial laser pulse polarizes the NV to the ms\textsubscript{ }=0 spin sublevel, after which the MW driving time $\tau$ is varied to achieve a varying NV\textsuperscript{-} spin population by coherently driving the spin. Finally the readout pulse reveals the NV\textsuperscript{-} spin population in the m\textsubscript{s}=0 or m\textsubscript{s}=$\pm$ 1 states.

\begin{figure}[!ht]
\centering
\includegraphics[width=\linewidth]{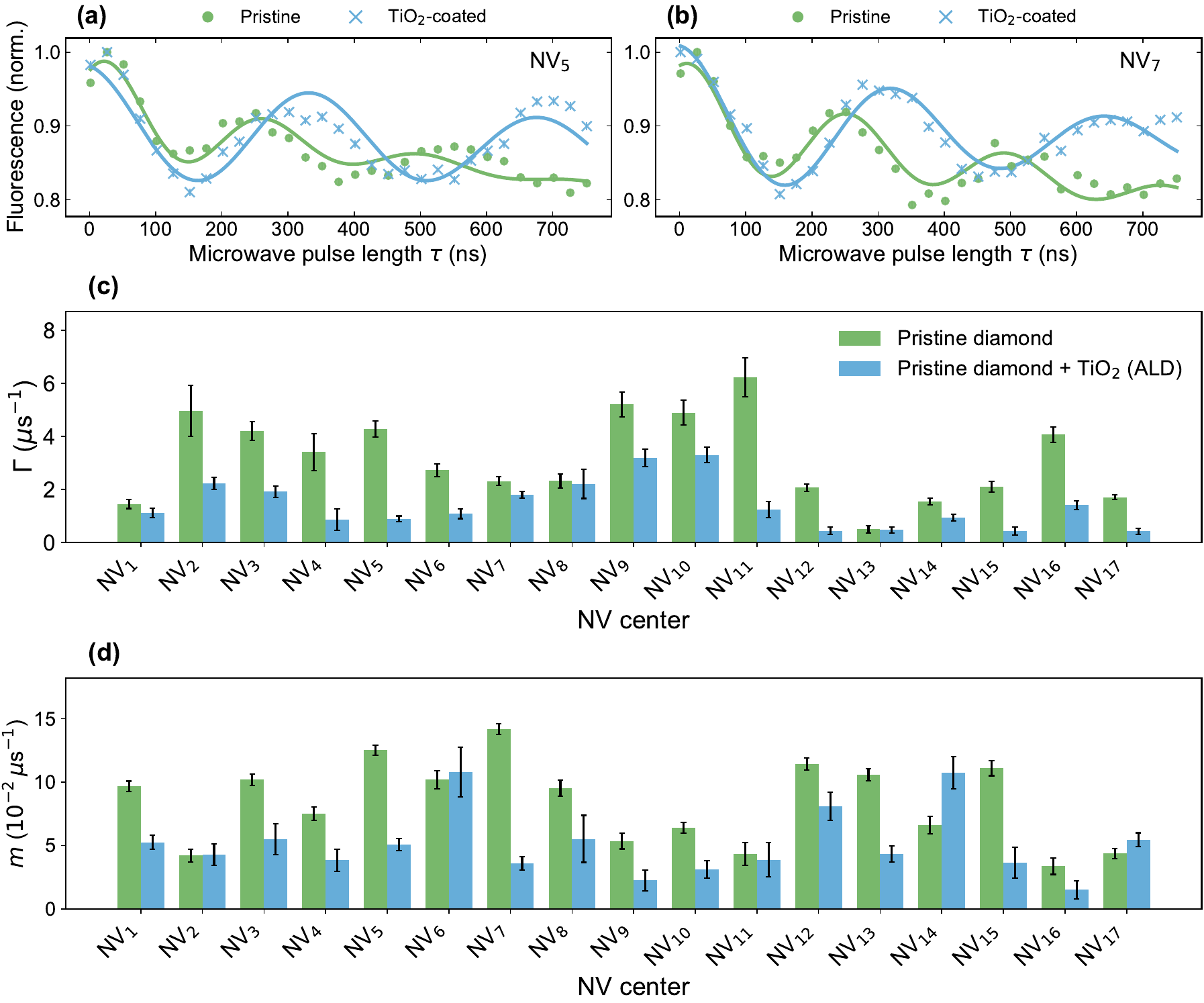}
\caption{\textbf{Improvement of Rabi oscillations of single NV centers in near-UHV condition} Driven Rabi oscillations for (a) NV\textsubscript{5} and (b) NV\textsubscript{7} in pristine diamond and in TiO\textsubscript{2} coated diamond. Statistics of (c) decay rate $\Gamma_{D}$ and (d) slope m in pristine and coated diamond conditions 
We observe a gradual decrease in the baseline for NV centers in pristine diamond which recovers after ALD coating. P= 3 x 10\textsuperscript{-9} mbar, T= 300 K, incident laser power before the objective = 130 $\mu$W. } 
\label{fig: rabi}

\end{figure}

We present the Rabi oscillation measurements before and after coating with TiO\textsubscript{2} for 2 NV centers (NV\textsubscript{5}  and NV\textsubscript{7}) in Figure \ref{fig: rabi} (a) and (b) respectively. The data is fit with an exponentially damped cosine model with a linear baseline term to account for systematic drifts, given by

\begin{equation}
  PL =  A*cos ( 2 \pi f_{rabi} * ({\Large \tau}-{\Large \tau_{0}})) * e^{-{\Large \tau}*\Gamma_{D}} - m*\tau + c 
\end{equation}

where the oscillation amplitude A, Rabi frequency f$_{rabi}$, phase $\tau_{0}$, cosine decay rate $\Gamma_{D}$ and constant PL offset c  are free parameters. As can be seen in Figure  \ref{fig: rabi} (a) and (b), the oscillations are more enhanced after the surface treatment. The fit yields a significant reduction in decay rate $\Gamma_{D}$, decreasing from 4.3 $\pm$ 0.3 $\mu s^{-1}$ to 0.9 $\pm$ 0.1 $\mu s^{-1}$ for NV\textsubscript{5} , and  2.3 $\pm$ 0.2 $\mu s^{-1}$  to 1.8 $\pm$ 0.1   $\mu s^{-1}$ for NV\textsubscript{7}, confirming this observation. The decay rate $\Gamma_{D}$  which governs the damping of the Rabi oscillation amplitude is strongly linked to the coherence of the NV spin. As a result of fluctuating charges and surface spin noise \cite{Lo2025EnhancementEngineering,Janitz2022DiamondCenters} the coherence times are significantly impacted near the surface \cite{Sangtawesin2019OriginsSpectroscopy}. The reduction of $\Gamma_{D}$ of the same NV center after the surface treatment therefore points to a significant reduction of the surface noise. To verify this observation, we examine the measured oscillations in all 17 NV centers. The extracted parameters as shown in  \ref{fig: rabi} (c) reveal a similar effect of reduced $\Gamma_{D}$, pointing to the improved ability to drive the NV spin without damping. In particular, the average $\Gamma_{D}$ reduces from 3.2 $\pm$ 0.4 $\mu s^{-1}$  to 1.4 $\pm$  0.2 $\mu s^{-1}$ averaged over 17 NV centers. For 3 of the analysed NV centers, the number of oscillations within the 750 ns measurement window was insufficient for a fully unconstrained fit (see SI Figure 8 for further details). %The oscillation frequency was therefore constrained to the value obtained by Fast Fourier Transform (FFT) of the data, while all other parameters including $\Gamma_{D}$ were kept free (see SI Figure 7). 
Although dedicated pulse sequences (such as Hahn echo sequence for T\textsubscript{2} measurement) and extended measurement windows are required to accurately determine coherence times, the systematic reduction in $\Gamma_{D}$ already provides substantial evidence for an improved ability to coherently drive the NV spin, attributed to the reduction in surface noise.

%it was not possible to reliably extract the parameters with the free fit procedure with the model mentioned before, due to having low number of oscillations within our measurement window of 750 ns, with oscillation frequency $<$ 2 MHz. Hence these data are treated separately by fixing the oscillation frequency obtained by FFT of the data as f\textsubscript{Rabi}, and then optimizing for amplitude A, phase $\tau_{0}$,  $\Gamma_{D}$  slope m, and PL offset c (see SI Figure 7 ). 

%The rest of the 14/17 NV centers are treated with the fitting model mentioned before. Notably, the reduction in $\Gamma_{D}$ is consistently observed  across all measured NV centers, resulting in the average decay rate changing from 3.1 $\pm$ 0.2 $\mu s^{-1}$  to 1.3 $\pm$  0.1 $\mu s^{-1}$ 
%averaged over 14 NV centers. The decay rate $\Gamma_{D}$  which governs the damping of the Rabi oscillation amplitude is strongly linked to the coherence times of the NV centers. 

Additionally, we find a repeating trend in the extracted values of the slope term $m$. For NV\textsubscript{5}, the value of $m$ changes from (12.5 $\pm$ 0.4) × 10\textsuperscript{-2} $\mu s^{-1}$ to (5.1 $\pm$ 0.5) × 10\textsuperscript{-2} $\mu s^{-1}$ and for NV\textsubscript{7} it changes from (14.2 $\pm$ 0.4) × 10\textsuperscript{-2} $\mu s^{-1}$ to (3.6 $\pm$ 0.5) × 10\textsuperscript{-2} $\mu s^{-1}$. As shown in Figure \ref{fig: rabi} (d),  this term consistently decreases upon coating, indicating that the baseline decay is suppressed after surface treatment.  From the averaged value of the slope over 17 NV centers, we find a change from (8.3 $\pm$ 0.8) × 10\textsuperscript{-2} $\mu s^{-1}$ to (5.1 $\pm$ 0.6) × 10\textsuperscript{-2} $\mu s^{-1}$. These results indicate that the baseline drift cannot be attributed to spurious experimental effects, but rather supports the picture of improved charge-state dynamics. In previous works \cite{Bluvstein2019IdentifyingCenters,Yuan2020ChargeDiamond}, the ionization of NV in the dark, i.e in the absence of the laser irradiation has been established. In the Rabi measurement protocol, as the MW on-time $\tau$ increases, the NV spends an increasing amount of time in the absence of the laser irradiation. The extracted values of $m$ indicate that this baseline decay occurs on a timescale approximately two orders of magnitude slower than the amplitude damping characterized by $\Gamma_{D}$, consistent with a gradual relaxation into the NV\textsuperscript{0} charge state at longer dark intervals. % Again, only the NV\textsuperscript{-} contributes to oscillating fluorescence as the spin population varies with varying MW drive time, while the NV\textsuperscript{0} contributes to background fluorescence \cite{Alsid2019PhotoluminescenceDiamond,CardosoBarbosa2023ImpactRelaxometry}.
The resulting fluorescence decay can therefore be attributed to an increasing likelihood of relaxing into the NV\textsuperscript{0} charge state at longer dark intervals, consistent with earlier reports\cite{Bluvstein2019IdentifyingCenters,Yuan2020ChargeDiamond}. The reduction in m together with the improvement in $\Gamma_{D}$, indicates enhanced charge-state stability in the dark after surface treatment. 
We therefore conclude that charge-state ionization dynamics that are rapidly exacerbated under UHV conditions, resulting in the slope decay in oscillations recored before the surface treatment. The surface treatment shifts the equilibrium towards more stable NV\textsuperscript{-} and mitigates this decay , therefore restoring the oscillations towards a behavior predicted by a conventional cosine damped model.

Through the comprehensive experimental evidence from the recorded fluorescence spectrum, pulsed ODMR and Rabi oscillations, we demonstrate the beneficial impact of the surface modification for stabilizing shallow NV centers in UHV. We attribute the observed effects of the surface modification to a change in the local electronic environment of the NV center, arising from reduced upward band bending and suppressed surface noise. Adsorbates on the surface have been shown to affect the stability of NV centers through a band-bending mechanism \cite{Hauf2011ChemicalDiamond,Grotz2012ChargeDiamond,Newell2016SurfaceDiamond,Neethirajan2023ControlledCenters}. At the diamond surface, accumulation of negative charges generates surface-induced electrostatic fields, causing upward band bending. As a result, the energy of the NV increases with proximity to the surface, while the Fermi level is constant from the surface to the bulk of the diamond. The charge transition level (CTL) of the NV center corresponds to the ground state of the NV\textsuperscript{-} state \cite{Weber2010QuantumDefects}. Near the surface, the CTL lies above the Fermi level, resulting in lower probability of the state being occupied by an electron, thus favouring the NV\textsuperscript{0} state. %Close to the surface, the CTL is above the Fermi level, favouring the NV\textsuperscript{0} state.  
 Moving deeper into the diamond, the CTL  drops below the Fermi level, thus favouring the NV\textsuperscript{-} inside the bulk of the diamond \cite{Hauf2011ChemicalDiamond}. In ambient conditions, adsorption of a thin water layer from the atmosphere onto the surface counteracts the upward bending and reduces the bending strength \cite{Hauf2011ChemicalDiamond}. This leads to the CTL crossing the Fermi level close to the surface, resulting in relatively stable NV\textsuperscript{-} near the surface. On the other hand, under UHV conditions, the adsorbed water layer may easily undergo changes, restoring the stronger band bending and the CTL crossing the Fermi level deeper inside the diamond. This leads to a higher probability of NV\textsuperscript{0} near the surface. Our results of charge-state stabilization can thus be explained by the TiO\textsubscript{2} layer passivating the charge traps and restoring the band bending to a state similar to ambient conditions. In addition, the passivation of surface charge traps by the TiO\textsubscript{2}
 coating is expected to reduce fluctuating charges and thus surface noise, consistent with the observed improvement in $\Gamma_{D}$. While not explored here, investigation of the effects of different layer thickness and surface morphology studies are expected to elucidate the detailed mechanism of the NV charge-state stabilization reported in this work. 

Lastly, based on the consistently observed improvement in charge-state stability and spin-state control, we evaluate the potential of the capping layer for precise and controlled nanoscale magnetometry using shallow NV centers. Similar oxide layers like Al\textsubscript{2}O\textsubscript{3} have already been used for surface functionalization and immobilization of biomolecules \cite{Liu2022SurfaceDiamond,Xie2022BiocompatibleSensor}.
The surface treatment used here extends beyond this by simultaneously serving as a controlled method for stabilizing shallow NV centers, as well as a basis for further surface treatments. With the possibility to easily remove and redeposit the coating, our results allow for a reconfigurable architecture compatible with biomolecules. The use of single NV centers enables high spatial resolution for both surface sensing as well as scanning NV operations. The tip-shaped nanostructures closely resemble conditions routinely used in a scanning NV platform, demonstrating the readiness of the method for application on combined AFM-NV scanning probe geometry. With optimized surface treatment, quantum magnetometers based on NV centers which are stable in a broad range of pressure conditions can then be realized. Therefore we envision that controlled surface treatment with atomic scale precision is a promising direction for enhancing performance of near-surface atomic scale quantum sensors.    
%To summarize, we perform spectroscopic measurements aimed to gain insight into the charge and spin dynamics of individual NV centers. With the analysis of the recorded fluorescence spectra, the pulsed ODMR and the Rabi oscillation behavior, we demonstrate that a controlled thin layer of TiO\textsubscript{2} counteracts the detrimental effects of the surface under UHV conditions. To our knowledge, this is the first demonstration of consistent charge stabilization under extreme conditions of UHV (P= 3 x 10 \textsuperscript{-9} mbar). Our results provide insight on charge-state dynamics in near-surface NV centers and have direct implications for future, improved NV magnetometers. 
Near-term experiments showing improvements in more complex sensing protocols such as Double Electron-Electron Resonance (DEER) spectroscopy will help establish controlled surface treatments as the routine approach for operation of shallow NV centers under extreme conditions.  

\section{Methods}

\subsection*{Sample preparation and measurements}
The primary diamond membrane sample was used for all the measurements involving NV dynamics. The electronic grade CVD-grown diamond sample from Element Six with (100) crystal orientation is implanted with 2.5, 5 and 10 keV \textsuperscript{15}N ions in separate regions. The sample is then annealed at 950 °C for 2 h to form the NV centers. The diamond surface was subsequently structured into nanopillar arrays using electron beam lithography and reactive-ion etching-inductively coupled plasma (RIE-ICP) recipes.  These pillars serve as waveguides for enhanced photon collection. Only the 5 keV implanted region is used for measurements reported here. Oxygen termination of the diamond surface was achieved through a treatment of
tri-acid mixture ( H\textsubscript{2}SO\textsubscript{4}:HCLO\textsubscript{4}:HNO\textsubscript{3} in 1:1:1 ratio) at 200 °C for 6 h. This oxygen termination serves as a base for the subsequent surface treatments. The surface is then cleaned using Piranha solution ( 3:1 H\textsubscript{2}SO\textsubscript{4}:H\textsubscript{2}O\textsubscript{2}) for 3 h to obtain a pristine diamond surface.  A gold wire is bonded across the sample for microwave delivery. Subsequently, the sample is transferred to the main measurement chamber (P = 3 x 10\textsuperscript{-9} mbar) after taking it through the loadlock chamber and preparation chamber at pressures P = 1 x 10\textsuperscript{-7} mbar and P = 1 x 10\textsuperscript{-9} mbar respectively without cracking the vacuum. At this stage, the measurements labeled pristine diamond are carried out. After measurements on the pristine diamond, the sample is taken out of the UHV measurement chamber and cleaned again with Piranha solution for 3 h. Then a layer of TiO\textsubscript{2} is deposited on the diamond using atomic layer deposition (ALD Cambridge Nanotech Savannah S100) with a Tetrakis(dimethylamino)titanium (TDMAT) precursor. The sample is transferred to the measurement chamber (following the same UHV transfer process mentioned above) and at this stage the measurements labelled "pristine diamond + TiO\textsubscript{2}" are carried out. All measurements on a particular NV center are carried out within 810 s (total measurement time). Laser power (130 $\mu$W) and acquisition times are kept constant across all NV centers for both pristine and ALD coated diamond for reliable comparison across different surface conditions. A Kratos Axis Ultra system with monochromatized Al K$\alpha$ was used for XPS measurements on the reference sample. The survey scans were acquired with a pass energy of 80 eV. A charge neutralizer was used for compensation of surface charging. The C 1s was calibrated to 284.8 eV (adventitious carbon)\cite{Biesinger2022AccessingReview}.

\begin{acknowledgement}
A.S. acknowledges support through her Emmy Noether grant from the Deutsche Forschungsgemeinschaft (DFG), project No. 504973613, the IQST young researcher grant, as well as the financial support by the Deutsche Forschungsgemeinschaft (DFG, German Research Foundation) through the Würzburg–Dresden Cluster of Excellence ctd.qmat – Complexity, Topology and Dynamics in Quantum Matter (EXC 2147, project-id 390858490). The authors also acknowledge discussions with Dr. Ruoming Peng, Sandip Maity and Dr. Ricardo Javier Pená Román. We acknowledge technical assistance from Wolfgang Stiepany, Isabel Pfander and Marko Memmler. 
\end{acknowledgement}

\bibliography{references}

\newpage

\begin{figure}
  \centering
\includegraphics[width=4.25 in,keepaspectratio]{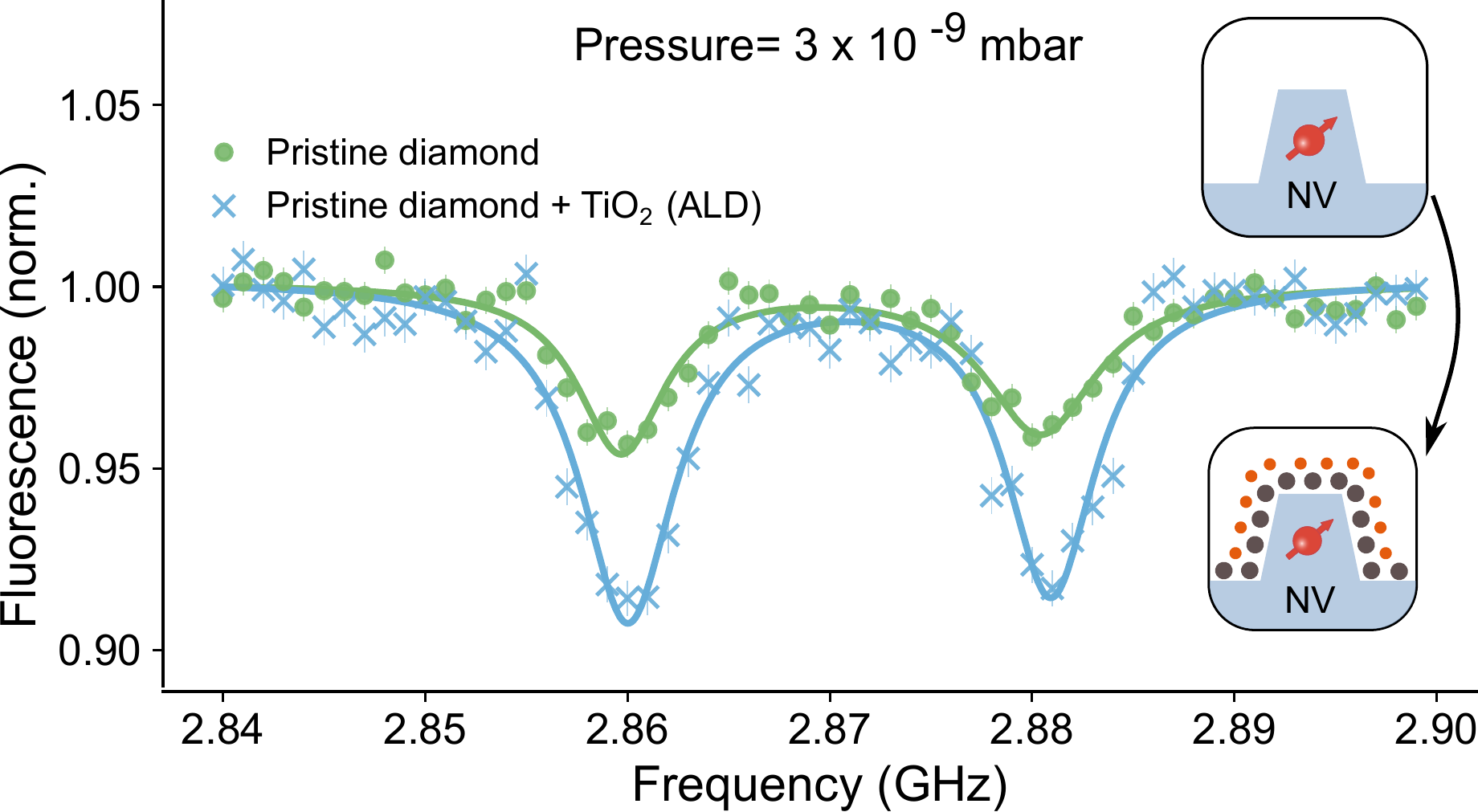}
\caption{For Table of Contents Only}
\end{figure}

\end{document}